# Image-free real-time classification of fast moving objects using 'learned' spatial light modulation and a single-pixel detector

Zibang Zhang[1], Xiang Li[1], Manhong Yao[1], Shujun Zheng[1], Guoan Zheng[2], and Jingang Zhong[1,*]

[1] *Department of Optoelectronic Engineering, Jinan University, Guangzhou 510632, China*
[2] *Biomedical Engineering, University of Connecticut, Storrs, CT 06269, USA*
*\*tzjg@jnu.edu.cn*

**Abstract:** Objects classification generally relies on image acquisition and analysis. Real-time classification of high-speed moving objects is challenging, as both high temporal resolution in image acquisition and low computational complexity in objects classification algorithms are required. Here we propose and experimentally demonstrate an approach for real-time moving objects classification without image acquisition. As objects classification algorithms rely on the feature information of objects, we propose to use spatial light modulation to acquire the feature information directly rather than performing image acquisition followed by features extraction. A convolutional neural network is designed and trained to learn the spatial features of the target objects. The trained network can generate structured patterns for spatial light modulation. Using the resulting structured patterns for spatial light modulation, the feature information of target objects can be compressively encoded into a short light intensity sequence. The resulting one-dimensional signal is collected by a single-pixel detector and fed to the convolutional neural network for objects classification. As experimentally demonstrated, the proposed approach can achieve accurate and real-time classification of fast moving objects. The proposed method has potential applications in the fields where fast moving objects classification in real time and for long duration is required.

## 1. Introduction

Objects classification systems constitute a deeply entrenched and omnipresent component of modern intelligent systems. Objects classification has found tremendous applications in various fields, such as traffic control [1], assembly-line industrial inspection [2], remote sensing [3], medical cytometry [4], and military [5]. Most available objects classification systems are based on image acquisition and analysis (also known as machine vision). Objects classification algorithms rely on the feature information of the target objects. Thus, image-based objects classification systems use an acquire-and-extract strategy. Specifically, the systems first employ photography to acquire image or images of the target objects and secondly deploy certain image analysis algorithm to extract the object features from the image or images for classification.

Although image-based objects classification systems have been explored for decades [6-8], high-speed moving objects classification in real time and for long duration is still a challenge. High-speed moving objects would result in dramatic motion blur in the images captured. Motion blur causes image quality degradation and, therefore, reduces classification accuracy. High-speed photography [9-11] and sophisticated image analysis algorithms [12-13] are approaches to deal with motion blur. However, it is challenging to apply high-speed photography in long-duration objects classification, as the data throughput of high-speed photograph is generally huge while data storage, bandwidth, and processing capacity are limited in practice. In addition, sophisticated image analysis algorithms are generally computationally exhausted and, therefore, difficult to apply for real-time classification.

Although only feature information is needed in objects classification, image-based objects classification systems commonly capture a complete image or even images for feature information acquisition, resulting in a huge amount of redundant data to be generated, transferred, and processed. In other words, the data efficiency of such an acquire-and-extract strategy is low.

Recently, S. Ota *et al.* proposed an image-free method for classification of high-speed moving blood cells, termed ghost cytometry [4]. Inspired by ghost imaging [14-15], this method uses a static and random pattern to modulate the illumination light field. As such, the feature information of high-speed moving cells is compressively encoded into a one-dimensional (1-D) light intensity signal. The resulting 1-D light signal is collected by using a single-pixel detector and fed to a support vector machine for cells classification. This method demonstrates that the image-free scheme enables higher data efficiency by avoiding redundant data acquisition. However, this method requires fluorescence labeling.

In this paper, we propose an image-free objects classification method which involves adaptive spatial light modulation and single-pixel detection. The proposed method uses an extract-and-acquire strategy. The feature information of the target objects is extracted by spatial light modulation and compressively modulated into a short 1-D light intensity signal. A single-pixel detector is used to collect the resulting signal for feature information acquisition. The collected 1-D signal is fed into a convolutional neural network (CNN) directly as input for objects classification. The CNN can not only achieve accurate classification, but also 'learn' the spatial features of the target objects and then generate structured patterns for spatial light modulation. The proposed method is both data- and computation-efficient, allowing real-time and long-duration classification of fast moving objects. Most recently, there are some other single-pixel imaging inspired methods proposed for static objects classification [16-18].

## 2. Principle

In recent years, the research of single-pixel imaging has demonstrated that it is an effective way for spatial information acquisition to combine spatial light modulation and single-pixel detection [19-22]. Spatial light modulation is referred to using a spatial light modulator to generate structured (time-varying and space-varying) patterns to modulate the light field that is used to illuminate objects (illumination light field) or to be measured by a photodetector (detection light field). Single-pixel detection is referred to measuring the resulting light signal with a spatially unresolvable photodetector (such as, photodiode or solar cell). Single-pixel imaging was initially proposed for image acquisition. Single-pixel imaging techniques use random [21] or orthogonal patterns [22] for spatial light modulation and reconstruct the object image through correlation [21], inverse orthogonal transformation [22], or compressive sensing [20].

Rather than acquiring a complete image by single-pixel imaging, the proposed method only acquires the feature information of the target objects only. To achieve spatial feature information acquisition, the core is to design some structured patterns, which are adapted to the target objects, for spatial light modulation. The intensity of the modulated light field is a metric for the similarity of the structured patterns and the target objects. The intensity can be measured by using a single-pixel detector, termed single-pixel measurement. Specifically, if the structured pattern and the target object under spatial light modulation have spatial features in common, the magnitude of the resulting single-pixel measurements would be large, and vice versa. Using a sequence of structured patterns for spatial light modulation allows one to have a corresponding sequence of single-pixel measurements. Every single classification of objects has a unique sequence of single-pixel measurements. With optimized structured patterns, the single-pixel measurement sequence allows objects of one classification to be accurately differentiated from another.

However, it is difficult to analytically express the object features. Thus, it is necessary to derive such features in an automatic and adaptive way. As promising research on deep learning has proven CNN is capable for automatic feature information extraction, we propose to use a CNN to generate structured patterns for spatial light modulation. Specifically, the convolution kernels of a trained CNN are used as the structured patterns. Although initially grayscale, the convolution kernels can be binarized using a certain algorithm without causing pronounced errors. The resulting binarized patterns can be displayed on a high-speed spatial light modulator, such as digital micro-mirror device (DMD), to ensure a high temporal resolution for fast moving objects. As will be demonstrated, the CNN can not only generate structured patterns for spatial light modulation, but also achieve objects classification by using the single-pixel measurement sequence as input. The whole objects classification process employs no image acquisition, ensuring high efficiency in data acquisition. Additionally, the finely designed CNN enables computationally efficient objects classification, allowing fast moving objects to be classified in real time.

The implementation of the proposed image-free method can be summarized as the following 5 steps:
1) CNN architecture design;
2) CNN training;
3) Extracting the convolution kernels from the trained CNN and conducting binarization;
4) Illuminating the target moving object with the binarized patterns and collecting the resulting light intensity with a single-pixel detector;
5) Feeding the CNN with the single-pixel measurement sequence as input and deriving the classification results from the output layer of the CNN.

## 3. Experiments

As a proof-of-concept, we use the proposed method for classification of fast moving handwritten digits. The experimental set-up is shown in Fig. 1. We generate structured illumination by using a DMD (Texas Instruments DMD 4100) as a spatial light modulator and 10-watt white LED as a light source. The DMD displays different structured patterns sequentially and the patterns are projected onto the target objects through Lens 1. The illumination area on the disk is 45 mm × 45 mm. A photodiode (Thorlabs PDA-100A2) is used as a single-pixel detector to collect the transmitted light. The 1-D light signal measured is digitalized using a data acquisition panel (National Instruments USB-6340). The CNN is trained and tested on a computer equipped with an Intel Core i3-7300 CPU and an NVidia RTX 2080Ti graphic card. We simulate fast moving objects by rotating a black disk with 8 handwritten digits uniformly carved. The carved digits are randomly chosen from the MNIST test set and printed on transparent films. The radius of the disk is 0.15 m. The disk is driven by a controllable motor whose rotation speed can be tuned by adjusting the input voltage. The structured patterns used for objects feature acquisition are generated by using the CNN shown in Fig. 2(a).

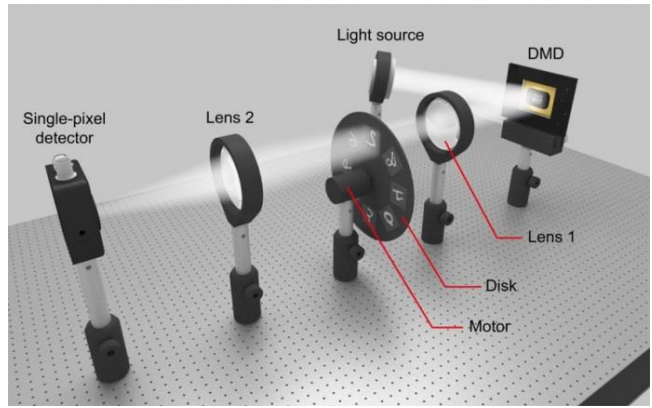

Fig. 1. Experimental set-up.

### 3.1 Architecture of proposed CNN

Figure 2(a) shows the architecture of the proposed CNN. The CNN consists of an image input layer, a convolutional layer, a deconvolutional layer, 3 fully connected layers, and an output layer.

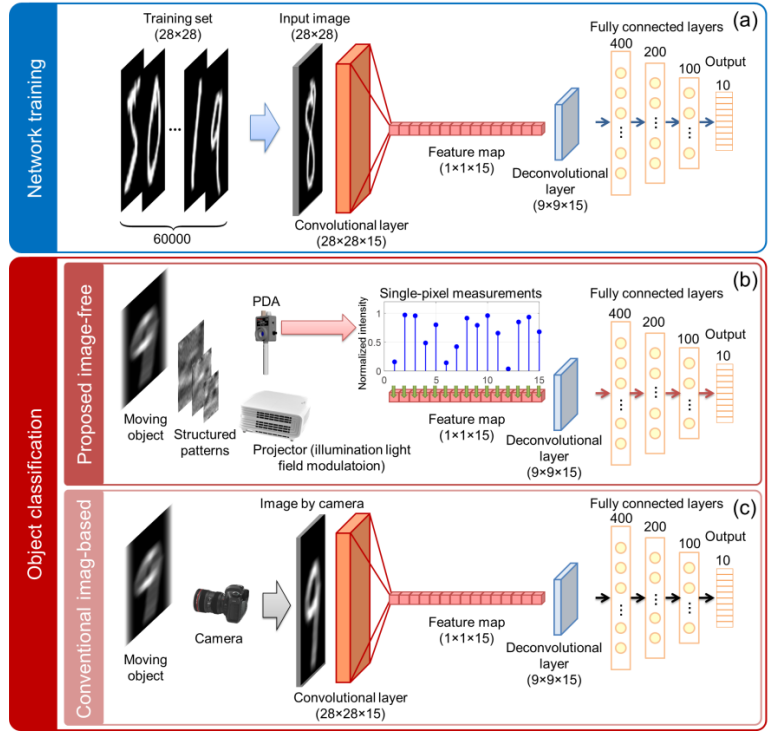

Fig. 2. Illustration of objects classification based on CNN. Architecture of proposed CNN (a) where image input layer accepts labeled training set images for network training. With such a designed network, our method allows objects classification to relieve image acquisition, which is done by feeding the resulting single-pixel measurements to the feature map (b). In comparison with conventional image-based counterpart (c), the proposed method achieves objects classification in an image-free manner.

As Fig. 2(a) shows, the image input is used to accept images as input for network training. The dimension of the image input layer is $28 \times 28$ so as to coincide with the size of images given by the MNIST handwritten digits database ($28 \times 28$ pixels) [23].

Following the image input layer, the convolution kernel layer is $28 \times 28 \times 15$ in size. In other words, the convolution kernel layer consists of 15 convolution kernels and the size of each kernel is $28 \times 28$. The size of the kernels is identical to the input images size. As a result, the convolution of the input image and each convolution kernel pattern is equivalent to an inner product. Such a design allows us to perform the convolution by a physical means, when the CNN is used for objects classification. Specifically, as Fig. 2(b) shows, we will use a spatial light modulator to illuminate target objects in the real world with convolution kernel patterns and use a single-pixel detector to measure the resulting light intensity. It has been demonstrated by single-pixel imaging techniques that the single-pixel measurements are the inner products of the target objects images and the convolution kernel patterns [20]. The single-pixel measurements will be fed to the feature map for proceeding objects classification by numerical computation. In comparison with objects classification by typical image-based scheme (Fig. 2(c)), the proposed method can achieve 'physical convolution' without acquiring the images of the target objects and doing the convolution through numerical computation.

The feature map in between the convolution kernel layer and the deconvolutional layer is an intermediate keeping the result of convolutions. It, therefore, consists no unknowns to be determined in the network training. Interestingly, we can feed single-pixel measurements, as they are equivalent to the convolutional results, to the feature map for proceeding objects classification, as Fig. 2(b) shows.

The deconvolutional layer is used for features upsampling. To increase nonlinearity of the network, the feature map of the deconvolutional layer is flattened and connected to the fully connected layers.

There are 3 fully connected layers following the deconvolutional layer. Each has 400, 200, 100 units, respectively.

In the output layer, the Softmax function is used to predict the final output.

### 3.2 Training set images preparation and CNN training

We use the training images provided by the database for CNN training. MNIST dataset has 60,000 training images and 10,000 testing images, respectively. The original images in database are upright and centered. However, objects in motion might have rotation and translation. In order to improve the robustness of the CNN for moving objects classification, we apply random translation (-11 to 11 pixels) and rotation (-15 to 15 degrees) to each training image. Figure 3 shows an example of 8 original images and their corresponding shifted and rotated version.

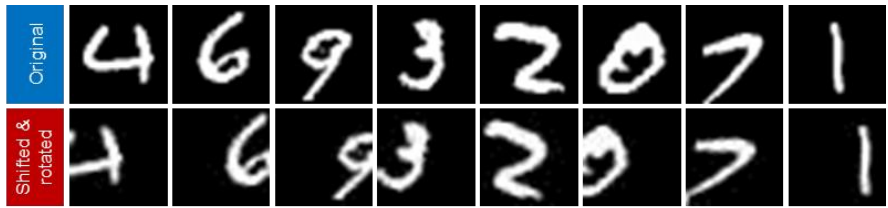

Fig. 3. Example of testing images with rotation and lateral shift. The first row shows 8 original handwritten digit training images and the second row shows the corresponding images with random lateral shift and rotation.

For network training, the network is initialized by truncated normal distribution with 0.1 standard deviation and 0 mean. All the bias terms are initialized with 0. Cross entropy loss function is adopted for optimizing. Estimated by the loss function, the error is back propagated through the network and the network's parameters are updated by the adaptive moment estimation (ADAM) optimization [24], which is a stochastic optimization method. We set a learning rate parameter of 0.001 and a mini-batch size of 128 images. The training process consumes 725.96 seconds on our computer.

### 3.3 CNN testing with static images input

To test the trained network, we use the testing images provided by the database through numerical simulation. The database includes 10,000 testing images. We feed the testing images to the image input layer. Among 10,000 testing images, 9,611 are correctly recognized and 389 are classified mistakenly. The overall accuracy in testing is 96.11%.

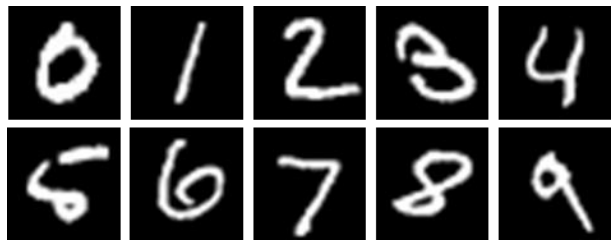

Fig. 4. Ten handwritten digit images used in network testing.

The trained CNN has a high overall accuracy. The confidence for every single testing digit is high as well. We present the output confidence for 10 among all 10,000 testing images in Table 1. The selected 10 testing images are shown in Fig. 4. As the Table 1 shows, the confidence of all 10 digits is close to 1, except that the confidence of digit '3' is only ~0.6. As is shown in Fig. 4, the digit '3' looks quite similar to digit '8', which leads to a relatively lower confidence in classification.

**Table 1. Output confidence of 10 among all 10,000 testing images in CNN testing**

| Digits | | Predictions | | | | | | | | |
|---|---|---|---|---|---|---|---|---|---|---|
| | | 0 | 1 | 2 | 3 | 4 | 5 | 6 | 7 | 8 | 9 |
| Labels | 0 | 0.9997 | 0.0000 | 0.0000 | 0.0000 | 0.0000 | 0.0000 | 0.0000 | 0.0000 | 0.0000 | 0.0002 |
| | 1 | 0.0000 | 0.9999 | 0.0000 | 0.0000 | 0.0000 | 0.0000 | 0.0000 | 0.0000 | 0.0000 | 0.0001 |
| | 2 | 0.0000 | 0.0000 | 1.0000 | 0.0000 | 0.0000 | 0.0000 | 0.0000 | 0.0000 | 0.0000 | 0.0000 |
| | 3 | 0.0044 | 0.0010 | 0.0994 | 0.5963 | 0.0041 | 0.0330 | 0.0026 | 0.0004 | 0.2571 | 0.0015 |
| | 4 | 0.0000 | 0.0000 | 0.0000 | 0.0000 | 0.9995 | 0.0000 | 0.0000 | 0.0001 | 0.0000 | 0.0004 |
| | 5 | 0.0000 | 0.0000 | 0.0000 | 0.0000 | 0.0000 | 1.0000 | 0.0000 | 0.0000 | 0.0000 | 0.0000 |
| | 6 | 0.0014 | 0.0000 | 0.0081 | 0.0001 | 0.0016 | 0.0366 | 0.9463 | 0.0000 | 0.0000 | 0.0000 |
| | 7 | 0.0000 | 0.0000 | 0.0000 | 0.0001 | 0.0000 | 0.0000 | 0.0000 | 0.9999 | 0.0000 | 0.0000 |
| | 8 | 0.0052 | 0.0036 | 0.0068 | 0.0013 | 0.0001 | 0.0003 | 0.0001 | 0.0001 | 0.9822 | 0.0003 |
| | 9 | 0.0000 | 0.0007 | 0.0000 | 0.0037 | 0.0006 | 0.0120 | 0.0000 | 0.0492 | 0.0001 | 0.9337 |

### 3.4 Illumination patterns extraction from CNN and patterns binarization

The trained CNN has 'learned' the features of handwritten digits. Such feature information is condensed in the convolution kernels. We extract these convolution kernels from the CNN and use them as illumination patterns. Figure 5 shows the 15 convolution kernels of the trained CNN.

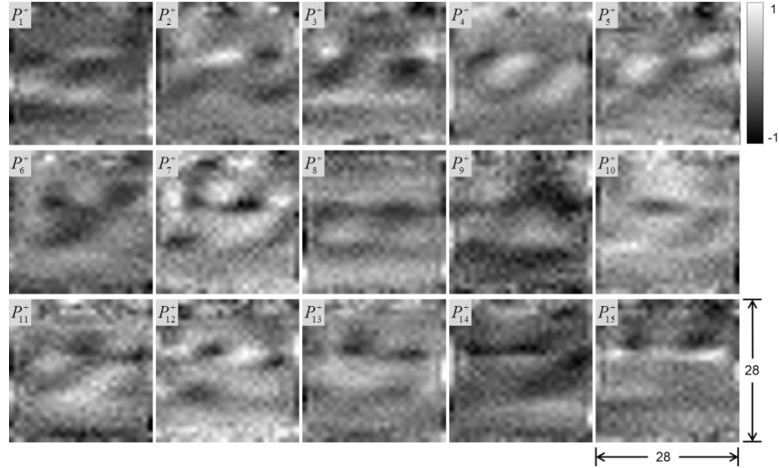

Fig. 5. Convolution kernels extracted from the trained CNN. The size of convolution kernels is $28 \times 28$ pixels.

As the figure shows, the elements in the convolution kernels range from -1 to 1. However, DMD as an intensity-only spatial light modulator can only generate patterns with non-negative values. Moreover, DMD can maximize its refreshing rate only when displaying binary patterns. To tackle this problem, we propose to binarize the convolution kernels and perform differential measurement. Specifically, we use a binarization strategy termed

'upsample-and-dither' [25] for patterns binarization. As Fig. 6 shows, the binarization strategy is two-step. First, the grayscale pattern to be binarized is upsampled by using the 'bicubic' interpolation algorithm with a specified ratio of $k$. In our experiment, we choose $k = 25$, resulting in each pattern using $700 \times 700$ micro mirrors to display. Second, the upsampled grayscale pattern is binarized by using the Floyd-Steinberg dithering algorithm [26]. The resulting patterns range from 0 to 1. Differential measurement is referred to illuminate the target objects with a binarized pattern $P^+$ followed by its inverse $P^-$ $(=1-P^+)$ and use the difference of the corresponding single-pixel measurements as one effective measurement.

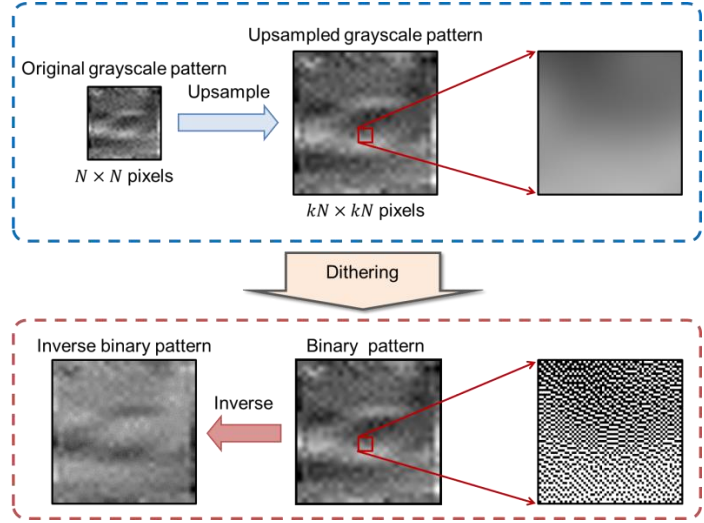

Fig. 6. Illustration of binary pattern generation.

### 3.5 Moving objects classification

We use the DMD to display the binary patterns and project the patterns onto the moving digits through Lens 1. The DMD operates at a refreshing rate of 17,850 Hz which is optimal in our experiment. A higher refreshing rate would result in dramatically lower classification accuracy. The PDA collects the resulting transmission light through Lens 2. By tuning the input motor voltage of the rotor, we can control the disk rotating at a different speed. We test the classification accuracy for different object moving speeds. We loop the 15 pairs of binary patterns for 400 times and collect 12,000 single-pixel measurements. The utilized data acquisition board operates at a data sampling rate of 500,000 Hz. We therefore take 28 (500,000 / 17,850) samples for each illumination pattern. For single-pixel measurement is referred to the average of 28 samples. For each classification, 30 single-pixel measurements are required. Thus, each classification consumes 1,680 bytes data (16 bits (2 bytes)/sample × 28 samples/measurement × 30 measurements/frame). The data throughput of the proposed method is 1,000,000 bytes (0.95 mega bytes) per second, which is much lower than high-speed photography. The raw data collected by the single-pixel detector are shown in Fig. 7.

Figure 7 shows the single-pixel measurements consist of a good many of ups and downs. As the digits on the disk are carved, the light can pass through the disk when a digit moves in the structured illumination area. Consequently, it results in single-pixel measurements of a high output voltage. On the contrary, when there is no digit in the structured illumination area, the light is blocked by the disk and the low output voltage is low. The classification results are shown in Table. 2. As the table shows, we also test the proposed system for static objects classification. As demonstrated by the results, the proposed method can successfully classify all moving objects when the linear speed of objects is not higher than 4.87 m/s. Considering the field-of-view (illumination area) is only 45 mm × 45 mm, each target object presents in the field-of-view is only ~0.01 seconds. Thus, we claim that the target objects are moving at a high speed. As the motor input voltage increases, the disk rotates faster and the classification accuracy decreases gradually. When the target objects move at a linear velocity of 8.24 m/s, the classification accuracy is still higher than 50%.

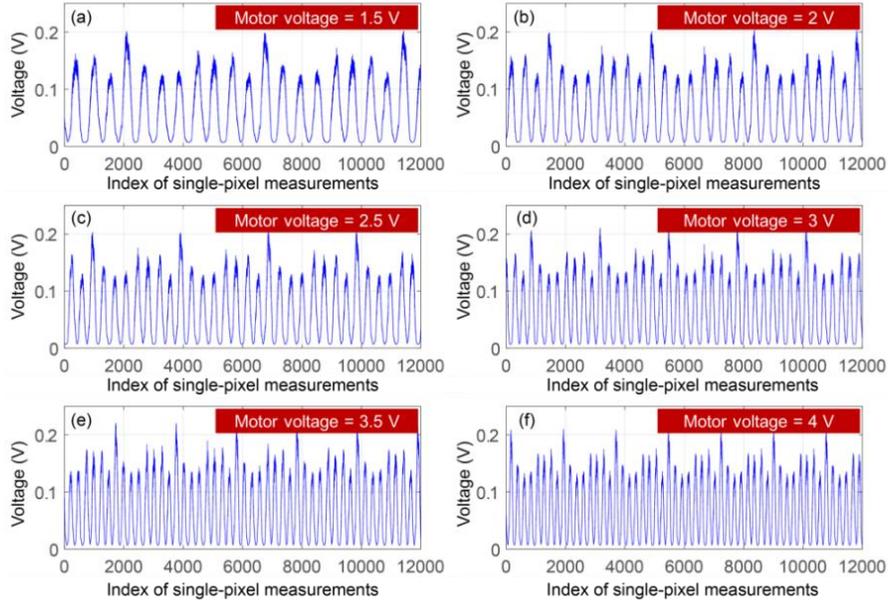

Fig. 7. Raw data captured by the single-pixel detector for target objects moving at different speeds controlled by using different motor input voltages.

The data acquisition time for each objects classification is 1.68 ms (30 patterns / 17,850 patterns/second). In other words, the temporal resolution of our method is 1.68 ms and the equivalent frame rate is 595 fps. The computational time for objects classification is 1.43 ms on average. As the computational time is shorter than the data acquisition time, the issue of data accumulation can be avoided. Thus, the proposed method enables fast moving objects classification in real time and for long duration.

Table 2. Classification results of moving handwritten digits at different moving speeds

| Motor voltage (V) | Angular speed (rad/s) | Linear speed (m/s) | Rotation speed (rpm) | Accuracy (%) |
|---|---|---|---|---|
| 0 | 0 | 0 | 0 | 100 |
| 1.5 | 24.05 | 3.61 | 229.2 | 100 |
| 2 | 32.49 | 4.87 | 310.2 | 96.43 |
| 2.5 | 37.74 | 5.66 | 360.6 | 96.88 |
| 3 | 48.34 | 7.25 | 462.0 | 78.57 |
| 3.5 | 54.94 | 8.24 | 525.0 | 85.11 |
| 4 | 63.31 | 9.50 | 604.8 | 62.96 |

We also compare the proposed method with high-speed photography. We use a high-speed camera (MIKROTRON, Eosens mini1) to capture images of the digits in motion at different moving speeds. We replace the single-pixel detector with the camera to capture two sets of images with a different frame rate, that is, 60 fps and 500 fps. Figure 8 presents the snapshots of digit '5' in motion at different speeds. The digit is moving so fast that it can hardly be recognized in the images for an exposure time of 1/60 seconds. The digit can be merely recognized by using an exposure time of 1/500 seconds. However, the camera can operate at 500 fps continuously for only 5.42 seconds.

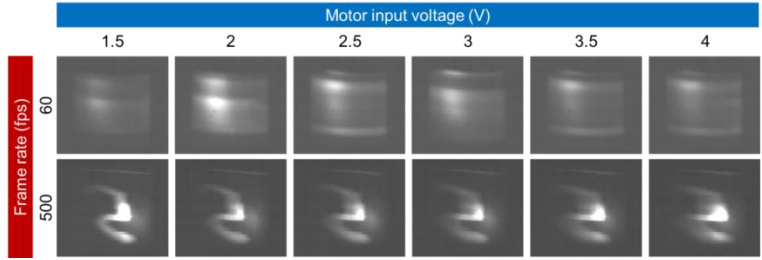

Fig. 8. Snapshots of digit '5' in motion at different speeds captured by using a 60-fps camera (top row) and a 500-fps camera (bottom row), respectively.

## 4. Discussion

The number of structured patterns varies with the number of categories of target objects and the similarity among categories. In our case, the number of categories of handwritten digits is 10, but the similarity between some handwritten digits is high, for example, 3 and 8, 1 and 7, 4 and 6, 3 and 9. Thus, we use 15 structured patterns (convolution kernels in the CNN) for objects classification. Generally speaking, the more categories of target objects, the more structured patterns are required. And the higher similarity among categories, the more structured patterns are required.

In our experiment, we use a DMD for spatial light modulation. However, the higher refreshing rate of the DMD results in a lower signal-to-noise ratio in single-pixel measurements, as the integral time for each single-pixel measurement will be shorter. It is reasonable to use a high-speed and high-flux spatial light modulator (for example, high-speed LED matrix [27,28]) so as to ensure a high signal-to-noise ratio even when the refreshing rate of the spatial light modulator is high.

Deep learning has been adopted in single-pixel imaging for reducing data acquisition time or improving image reconstruction quality [29-32]. Our method adopts deep learning to achieve adaptive spatial light modulation patterns generation and objects classification using single-pixel measurements, which might generate a new insight for spatial light modulation patterns optimization by using deep learning.

## 5. Conclusion

We report an image-free and real-time classification method for high-speed moving objects. The reported method is based on spatial light modulation technology, single-pixel detection, and deep learning. As experimentally demonstrated, the method can successfully classify objects moving as fast as 4.87 m/s in a field-of-view of 45 mm $\times$ 45 mm. The temporal resolution achieved is 1.68 ms and each classification acquires 1,680 bytes data and 1.43 ms for computation, which enables real-time and long-duration objects classification. As single-pixel detectors have the advantage of sensing at nonvisible wavebands, the reported method might potentially be used for hidden objects classification.

**Funding**

National Science Foundation of China (NSFC) (61905098 and 61875074) and Fundamental Research Funds for the Central Universities (11618307).

**Disclosures**

The authors declare no conflicts of interest.

**References**

1. D. Cireşan, U. Meier, J. Masci, and J. Schmidhuber, "Multi-column deep neural network for traffic sign classification," Neural Networks **32**, 333-338 (2012).
2. R. Zhao, R. Yan, Z. Chen, K. Mao, P. Wang, and R. X. Gao, "Deep learning and its applications to machine health monitoring," Mech. Syst. Signal Pr. **115**(15), 213-237 (2019).
3. X. Lu, X. Zheng, and Y. Yuan, "Remote Sensing Scene Classification by Unsupervised Representation Learning," IEEE T. Geosci. Remote **55**(9), 5148-5157 (2017).
4. S. Ota, R. Horisaki, Y. Kawamura, M. Ugawa1, I. Sato, K. Hashimoto, R. Kamesawa, K. Setoyama1, S. Yamaguchi, K. Fujiu, K. Waki, and H. Noji, "Ghost cytometry," Science **360**(6394), 1246-1251 (2018).
5. J. Neulist and W. Armbruster, "Segmentation, classification, and pose estimation of military vehicles in low resolution laser radar images," P. Soc. Photo-Opt. Ins. **5791**, 218-225 (2005).
6. O. Russakovsky, J. Deng, H. Su, J. Krause, S. Satheesh, S. Ma, Z. Huang, A. Karpathy, A. Khosla, M. Bernstein, A. C. Berg, and F. F. Li, "Imagenet large scale visual recognition challenge," Int. J. Comput. Vision **115**(3), 211-252 (2015).
7. A. Krizhevsky, I. Sutskever, and G. E. Hinton, "Imagenet classification with deep convolutional neural networks," Adv. Neur. In. **25**, 1097-1105 (2012).
8. A. Andreopoulos and J. K. Tsotsos, "50 years of object recognition: Directions forward," Comput. Vis. Image Und. **117**(8), 827-891 (2013).
9. M. Vollmer and K. P. Möllmann, "High speed and slow motion: the technology of modern high speed cameras," Phys. Educ. **46**(2), 191 (2011).
10. A. Veeraraghavan, D. Reddy, and R. Raskar, "Coded strobing photography: Compressive sensing of high speed periodic videos," IEEE T. Pattern Anal. **33**(4), 671-686 (2010).
11. D. Reddy, A. Veeraraghavan, and R. Chellappa, "P2C2: Programmable pixel compressive camera for high speed imaging," in Proceedings of IEEE Conference on Computer Vision and Pattern Recognition (IEEE, 2011), pp. 329-336.
12. L. Xu and J. Jia, "Two-phase kernel estimation for robust motion deblurring," in European conference on computer vision (2010), pp. 157-170.
13. J. Sun, W. Cao, Z. Xu, and J. Ponce, "Learning a convolutional neural network for non-uniform motion blur removal," in Proceedings of IEEE Conference on Computer Vision and Pattern Recognition (IEEE, 2015), pp. 769-777.
14. A. Gatti, E. Brambilla, M. Bache, and L. A. Lugiato, "Ghost imaging with thermal light: comparing entanglement and classical correlation," Phys. Rev. Lett. **93**(9), 093602 (2004).
15. J. H. Shapiro, "Computational ghost imaging," Phys. Rev. A **78**(6), 061802 (2008).
16. H. Chen, J. Shi, X. Liu, Z. Niu, and G. Zeng, "Single-pixel non-imaging object recognition by means of Fourier spectrum acquisition," Opt. Commun. **413**, 269-275 (2018).
17. S. Jiao, J. Feng, Y. Gao, T. Lei, Z. Xie, and X. Yuan, "Optical machine learning with incoherent light and a single-pixel detector," Opt. Lett. **44**(21), 5186-5189 (2019).
18. F. Wang, H. Wang, H. Wang, G. Li, and G. Situ, "Learning from simulation: An end-to-end deep-learning approach for computational ghost imaging," Opt. Express **27**(18), 25560-25572 (2019).
19. M. P. Edgar, G. M. Gibson, and M. J. Padgett, "Principles and prospects for single-pixel imaging," Nat. photonics **13**(1), 13-20 (2019).
20. M. F. Duarte, M. A. Davenport, D. Takhar, J. N. Laska, T. Sun, K. F. Kelly, and R. G. Baraniuk, "Single-pixel imaging via compressive sampling," IEEE Signal Proc. Mag. **25**(2), 83-91 (2008).
21. B. Sun, M. P. Edgar, R. Bowman, L. E. Vittert, S. Welsh, A. Bowman, and M. J. Padgett, "3D computational imaging with single-pixel detectors," Science **340**(6134), 844-847 (2013).
22. Z. Zhang, X. Wang, G. Zheng, and J. Zhong, "Hadamard single-pixel imaging versus Fourier single-pixel imaging," Opt. Express **25**(16), 19619-19639 (2017).
23. Y. LeCun, C. Cortes, and C. J. C. Burges, "THE MNIST DATABASE of handwritten digits," http://yann.lecun.com/exdb/mnist/.
24. D. P. Kingma and J. Ba, "Adam: A method for stochastic optimization," https://arxiv.org/abs/1412.6980 (2014).
25. Z. Zhang, X. Wang, G. Zheng, and J. Zhong, "Fast Fourier single-pixel imaging via binary illumination," Sci. Rep. **7**(1), 12029 (2017).
26. R. Floyd and L. Steinberg, "An adaptive algorithm for spatial grey scale," Proc. Soc. Inf. Display, **17**, pp. 75-77 (1976).
27. Z. Xu, W. Chen, J. Penuelas, M. Padgett, and M. Sun, "1000 fps computational ghost imaging using LED-based structured illumination," Opt. Express **26**(3), 2427-2434 (2018).
28. E. Salvador-Balaguer, P. Latorre-Carmona, C. Chabert, F. Pla, J. Lancis, and E. Tajahuerce, "Low-cost single-pixel 3D imaging by using an LED array," Opt. Express **26**(12), 15623-15631 (2018).
29. C. F. Higham, R. Murray-Smith, M. J. Padgett, and M. P. Edgar, "Deep learning for real-time single-pixel video," Sci. Rep.-UK **8**(1), 2369 (2018).
30. M. Lyu, W. Wang, H. Wang, H. Wang, G. Li, N. Chen, and G. Situ, "Deep-learning-based ghost imaging." Sci. Rep.-UK **7**(1), 17865 (2017).
31. T. Shimobaba, Y. Endo, T. Nishitsuji, T. Takahashi, Y. Nagahama, S. Hasegawa, M. Sano, R. Hirayama, T. Kakue, A. Shiraki, and T. Ito, "Computational ghost imaging using deep learning," Opt. Commun. **413**, 147-151 (2018).

32. Y. He, G. Wang, G. Dong, S. Zhu, H. Chen, A. Zhang, and Z. Xu, "Ghost Imaging Based on Deep Learning," Sci. Rep.-UK **8**(1), 6469 (2018).